# Thermoelectric signals of state transition in polycrystalline $SmB_6$

Z. J. Yue[1*], Q. J. Chen[2], and X. L. Wang[1]

*1. Institute for Superconducting and Electronic Materials, Faculty of Engineering, University of Wollongong, North Wollongong, NSW 2500, Australia*

*2. School of Physics and Electronics, Hunan University, Changsha 410082, China*

Email: zy709@uowmail.edu.au

Topological Kondo insulator $SmB_6$ has attracted quite a lot of attentions from condensed matter physics community. A number of unique electronic properties, including low-temperature resistivity anomaly, 1D electronic transport and 2D Fermi surfaces have been observed in $SmB_6$. Here, we report on thermoelectric transport properties of polycrystalline $SmB_6$ over a broad temperature from 300 K to 2 K. An anomalous transition in the temperature-dependent Seebeck coefficient $S$ from $S(T) \propto T^{-1}$ to $S(T) \propto T$ was observed around 12 K. Such a transition demonstrates a transition of conductivity from 3D metallic bulk states to 2D metallic surface states with insulating bulk states. Our results suggest that the thermotransport measurements could be used for the characterization of state transition in topological insulators.

Key words: Topological Kondo insulators, $SmB_6$, Seebeck effects, State transition.

PACS numbers: 72.20.Pa, 73.20.At

Topological insulators are quantum matters with topologically protected metallic surface states and insulating bulk states.[1] Fanscinating electronic properties, including the quantum spin Hall effects, exotic Majorana particles, and topological magnetoelectric effects have been observed in topological insulators.[2] With these novel properties, they hold high promising for next generation spintronics and quantum computing. Topological Kondo insulator $SmB_6$ that is a well-known strong correlated and heavy fermion system has recently attracted quite a lot of attentions from this field.[3] Various experimental measurements have been utilized to investigate the topological surface states in $SmB_6$ single crystals, which include magneto-transport, angle-resolved photoemission spectroscopy (ARPES), torque magnetometry and quantum oscillation.[3,4,5] Many unique electronic properties, including turely bulk-insulating, low-temperature resistivity anomaly, possible nematic phase, 1D edge state transport, and 2D unconventional Fermi surfaces have been observed in $SmB_6$.[3,4,6,7]

On the other hand, most discovered topological insulator materials, including $Bi_2Se_3$, $Bi_2Te_3$ and $Sb_2Te_3$, demonstrate high bulk conductivity due to defects induced bulk carriers.[8] In these topological insulators, the surface transport can be masked by highly conductive bulk states.[9] In contrast, in the $SmB_6$, the surface transport is dominant in low temperature due to truely insulating bulk states. There are large amounts of grain boundaries in polycrystalline $SmB_6$. These boundaries could form 2D surface states and conductivity channels in low temprature. Here, we report on thermoelectric transport measurements in polycrystalline $SmB_6$ and demonstrate a state transition from Seebeck effect. The Seebeck coefficient displays an anomalous transition from $S(T) \propto T^1$ to $S(T) \propto T$ around 12 K. Such a transition suggests a transition of conductivity from 3D metallic bulk states to 2D metallic surface states with insulating bulk states. In addition, our results provide evidences that topological surface states exist and is robust to large amounts of boundaries and defects in the polycrystalline $SmB_6$.

Polycrystalline samples of $SmB_6$ were fabricated in a vacuum furnace at a high temperature of 900 ˚C and a high pressure of 7 bar. The $SmB_6$ samples are in the shape of bars with typical dimensions of about $2 \times 2 \times 10$ mm$^3$. The Seebeck coefficient, electrical resistivity, and thermal conductivity were measured using a Quantum Design 14 T Physical Properties Measurement System (PPMS). The four-probe mode was employed in thermoelectric transport measurements between 2 K and 300 K. The applied AC bias current was 2 mA.

Figure 1(a) shows the temperature dependent resistivity of polycrystalline $SmB_6$ in the range from 2K to 300 K. Above 40 K, the bulk Kondo gap is closed due to thermal energy excitations, and the $SmB_6$ behaves like a metal. As the temperature decreases, the resistivity increases by several orders of magnitude due to the opening of a Kondo gap. Below 3.5 K, the resistivity approaches a constant value of ~ 0.017 $\Omega \cdot$m, which originates from the surface states of $SmB_6$.[6] In Figure 1(b), the Hall resistance is plotted as a function of magnetic fields at different temperatures. The negative sign in the Hall effect demonstrates that the carriers are electrons at low temperatures. The calculated carrier density in polycrystalline $SmB_6$ are about $3.25 \times 10^{18}$ $cm^{-3}$ and 1.5 $cm^2/V{\cdot}s$, respectively. Compared to $SmB_6$ single crystals, the mobility is quite low.[10] This is attributable to very high density of boundaries and defects in the polycrystalline $SmB_6$. Fig. 1(c) and (d) presents the magnetization as a function of temperature and magnetic fields, respectively. The polycrystalline $SmB_6$ demonstrates paramagnetic behaviour at high temperature. The low temperature Curie tail maybe related to the recent observed surface ferromagnetic domain walls.[7]

Figure 2(a) shows the total thermal conductivity, $\kappa$, decreases with the temperature decreasing. A subtle transition of the $\kappa$ happens at ~ 40 K which corresponds to an opening of the Kondo gap. As the Kondo gap opens, the contribution to the $\kappa$ from electrons decreases. The thermal conductivity is the sum of two contributions of electrons and phonons, which can be expressed as $\kappa = \kappa_e + \kappa_p$. At low temperatures, electron-phonon scattering increases, and the $\kappa$ is dominated by phonons. In addition, the reduced dimensions from 3D bulk states to 2D surface states also reduces the phonon contribution. At high temperature, large $\kappa$ stems from high density of bulk electrons with strong correlated interactions.

Fig. 2(b) shows the Seebeck coefficient, $S$, as a function of the temperature. A dramatically low-temperature peak of $S$ was seen at ~ 12 K. The low temperature value of the $S$ peak is -324 $\mu VK^{-1}$. Below 3.5 K, $S$ is linearly dependent on temperature. We find that the $S$ as a function of the temperature in $SmB_6$ represents a 3D bulk value at higher temperature with the relation $S(T) \propto T^{-1}$. With the temperature decreasing, the relation changes to $S(T) \propto T$ below a critical temperature. In this temperature range, the dominant carriers are electrons from 2D surface states.

Above 12 K, the Seebeck coefficient $S$ arises from the electronic contribution in the 3D $SmB_6$ bulk. Based on the Boltzman equation and the Mott relation, the Seebeck coefficient is described as,[11]

$$\boldsymbol{S} = -\frac{1}{eT}\frac{\int L(\epsilon)(\epsilon-u)\,(-\frac{\partial f}{\partial \epsilon})d\epsilon}{\int L(\epsilon)\,(-\frac{\partial f}{\partial \epsilon})d\epsilon} \qquad (1)$$

Here, $\boldsymbol{L(\epsilon) = \rho_c(\epsilon) v_c(\epsilon)^2 \tau_c(\epsilon)}$, where $\boldsymbol{\rho_c(\epsilon)}$ is density of states, $\boldsymbol{v_c(\epsilon)}$ is the electron velocity, $\boldsymbol{\tau_c(\epsilon)}$ is the electron relaxation time, and *u* is the chemical potential.

The formula yields the well known Seebeck coefficient for semiconductors,

$$\boldsymbol{S(T) \approx -\frac{k_B}{e}\frac{E_{c,v}-u}{k_B T}} \tag{2}$$

here $\boldsymbol{E_{c,v}}$ is the gap edge position of the conduction or valence band, and $k_B$ is Boltzmann's constant. Assuming that $\boldsymbol{\frac{du}{dT}}$ is negligible, *S* is proportional to $\boldsymbol{-\frac{1}{k_B T}}$ at high temperature.

Hence, $\boldsymbol{S(T) \propto -\frac{1}{T}}$ (3)

*S* is sensitive to the characteristics of the electronic structure. According to ARPES, the band structures of $SmB_6$ display a Kondo gap of about 15 meV at the Fermi level and surface states below 15 K.[12] The 2D metallic surface states also emerge below 15 K. Below $T$ = 12 K, the Seebeck coefficient *S* arises from the electronic contribution in the metallic surface states. R. Takahashi *et al.* have theoretically studied thermoelectric transport in 3D topological insulators by the Boltzmann equation. In a 2D topological insulator, the transport matrix for describing thermoelectric transport is,[13]

$$\begin{pmatrix} j/q \\ w \end{pmatrix} = \begin{pmatrix} \boldsymbol{L_0} & \boldsymbol{L_1} \\ \boldsymbol{L_1} & \boldsymbol{L_2} \end{pmatrix} \begin{pmatrix} -\frac{du}{dx} \\ -\frac{1du}{Tdx} \end{pmatrix} \tag{4}$$

Here, *j* is the electric current induced by electric field and *w* is the thermal current induced by thermal gradient. *q* is charge, and *u* is the chemical potential.

From the matrix elements, the thermoelectric parameters can be expressed as,

$$\boldsymbol{\sigma = e^2 L_0} \tag{5}$$

$$\boldsymbol{S = -\frac{1}{eT}\frac{L_1}{L_0}} \tag{6}$$

$$\boldsymbol{k_e = \frac{1}{T}\frac{L_0 L_2 - L^2{}_1}{L_0}} \tag{7}$$

$$\boldsymbol{ZT = \frac{L^2{}_1}{L_0 L_2 - L^2{}_1 + k_L T L_0}} \tag{8}$$

Here, the $\boldsymbol{k_e}$ and $\boldsymbol{k_L}$ is the thermal conductivity from electrons and phonons, respectively, $\boldsymbol{\sigma}$ is the electrical conductivity, and *ZT* is the thermoelectric figure of merit. In this theory, $\boldsymbol{k_L}$ is a constant. Considering a thin slab of 3D topological insulator with a small thickness *d*, so that the bulk is treated as 2D, from the Boltzman transport equation, the surface state transport can be expressed as,

$$L_v = c(k_B\mathrm{T})^v \int_{-\tilde{\Delta}}^{0} \frac{(x-\tilde{u})_0 e^{-x+\tilde{u}}}{(e^{-x+\tilde{u}}+1)^2} dx \quad (9)$$

$$c = \frac{1}{\pi L_y} \frac{\hbar v^2}{n_i V^2(0)} \quad (10)$$

$$\tilde{u} = \frac{u}{k_B T} \text{ and } \tilde{\Delta} = \frac{\Delta}{k_B T} \quad (11)$$

Here $v$ is the Dirac velocity, $V$ is the average impurity potential, and $n_i$ is the density of impurities.

Hence,

$$S = -\frac{1}{eT}\frac{L_1}{L_0} \quad (12)$$

$$S(T) = -\frac{k_B}{e} \frac{\int_{-\tilde{\Delta}}^{0} \frac{(x-\tilde{u})e^{-x+\tilde{u}}}{(e^{-x+\tilde{u}}+1)^2} dx}{\int_{-\tilde{\Delta}}^{0} \frac{e^{-x+\tilde{u}}}{(e^{-x+\tilde{u}}+1)^2} dx} \quad (13)$$

$$S(T) \propto -T \quad (14)$$

S. P. Chao *et al.* have calculated the thermoelectric transport in the surface states of topological insulators in the presence of randomly distributed impurities.[14] They deduced the generalized Mott formula for $S$ when the temperature is close to zero, $T \longrightarrow 0$,

$$S_{ij} = -\frac{\pi^2 {k_B}^2 T}{3e} \sum_m [\sigma^{-1}]_{im} [\partial\sigma/\partial u]_{mj} \quad (15)$$

Here, $\boldsymbol{\sigma}$ is the electrical conductivity,

When the impurity potential $V \neq 0$ and the chemical potential $u \approx 0$, $S \propto -\frac{k_B T}{u}$.

For a clean surface state with $V \approx 0$, the density of impurities $n \approx 0$,

$$S_{xx} \approx -\frac{\pi^2 k_B}{3e} \left(\frac{1}{u/k_B T} - \frac{\partial u \Im\Sigma(u)}{\Im\Sigma(u)/k_B T}\right) \quad (16)$$

The first term in Eq. (4) is the dominant thermopower in large $\mu$ or $u$, so

$$S \approx -\frac{\pi^2 k_B^2 T}{3ue} \quad (17)$$

Hence, overall, as $\mathrm{T} \longrightarrow 0$, $S(T) \propto -T$. The final Seebeck becomes

$$S = -\frac{1}{eT}\frac{L_1}{L_0} \quad (18)$$

As shown in Fig. 2(b), polycrystalline $SmB_6$ exhibits a low-temperature $S$ anomaly. Below 12 K, $|S|$ decreases linearly with temperature and resembles values for metal-like transport. High-resolution ARPES identified that in-gap low-lying states form electron-like Fermi surface pockets within a 4 meV window of the Fermi level.[5] They disappear above 15 K, in correspondence with the complete disappearance of the 2D conductivity channels. Hence, the $S$ anomaly below 12 K in polycrystalline $SmB_6$ is attributable to the 2D conductivity channels.

The efficiency of thermoelectric materials for cooling or power generation is described by the thermoelectric figure of merit ZT

$$ZT = \frac{S^2\sigma}{\kappa}T \tag{19}$$

where $\sigma$, $S$, and $\kappa$ are the electrical conductivity, the Seebeck coefficient, and the thermal conductivity respectively. A larger value of $ZT$ means a higher energy conversion efficiency. The figure of merit can be improved through increase $S$ and $\sigma$, while decrease $\kappa$. The resulting $ZT$ in polycrystalline $SmB_6$ is shown in Fig. 3. At high temperature, the 3D metallic bulk transport is dominant. The $ZT$ increases with decreasing temperature and reaches a maximum value of up to 0.0073 at 37 K. Below 37 K, the $ZT$ monotonously decreases down to near zero. The topological surface states provide a route to optimize the figure of merit $ZT$ and realize high performance thermoelectric devices.[15,16] The $ZT$ can be enhanced in 3D superlattices of topological insulator thin films and topological insulators with high densities of holes.[17] The $ZT$ can also be tuned by inducing a gap in the surface states through external magnetic fields and hybridization.[16,18] Particularly, in 3D topological Anderson insulators, large $ZT$ can be achieved via line dislocation engineering.[19] These makes 3D topological insulators strong candidates for applications in thermoelectric devices.

In summary, we report the thermoelectric transport in the polycrystalline topological Kondo insulator $SmB_6$. We find that the robust surface states survive in polycrystalline material with a large amount of non-magnetic impurities and disorder. The temperature dependent Seebeck coefficient demonstrates an anomalous transition from $S(T) \propto -1/T$ to $S(T) \propto -T$ at 12 K, where surface states start dominating the conductivity. This anomalous transition demonstrates a transformation from 3D metallic bulk states to completely 2D metallic surface states. Our results provide a new way to detect surface states in topological insulators.

Acknowledgements

This work is partially supported by the Australian Research Council under a Discovery project (ARC Discovery, DP130102956).

[a)]Author to whom correspondence should be addressed. Electronic mail: xiaolin@uow.edu.au.

## References

[1] M. Z. Hasan and C. L. Kane, Reviews of Modern Physics **82** (4), 3045 (2010).

[2] X.-L. Qi and S.-C. Zhang, Reviews of Modern Physics **83** (4), 1057 (2011).

[3] G. Li, Z. Xiang, F. Yu, T. Asaba, B. Lawson, P. Cai, C. Tinsman, A. Berkley, S. Wolgast, Y. S. Eo, D.-J. Kim, C. Kurdak, J. W. Allen, K. Sun, X. H. Chen, Y. Y. Wang, Z. Fisk, and L. Li, Science **346** (6214), 1208 (2014); B. S. Tan, Y.-T. Hsu, B. Zeng, M. C. Hatnean, N. Harrison, Z. Zhu, M. Hartstein, M. Kiourlappou, A. Srivastava, M. D. Johannes, T. P. Murphy, J.-H. Park, L. Balicas, G. G. Lonzarich, G. Balakrishnan, and S. E. Sebastian, Science (2015).

[4] Z. Yue, X. Wang, D. Wang, J. Wang, D. Culcer, and S. Dou, Journal of the Physical Society of Japan **84** (4), 044717 (2015).

[5] M. Neupane, N. Alidoust, S. Y. Xu, T. Kondo, Y. Ishida, D. J. Kim, C. Liu, I. Belopolski, Y. J. Jo, T. R. Chang, H. T. Jeng, T. Durakiewicz, L. Balicas, H. Lin, A. Bansil, S. Shin, Z. Fisk, and M. Z. Hasan, Nat Commun **4** (2013).

[6] D. J. Kim, J. Xia, and Z. Fisk, Nat Mater **13** (5), 466 (2014).

[7] Y. Nakajima, P. Syers, X. Wang, R. Wang, and J. Paglione, Nat Phys **advance online publication** (2015).

[8] Y. Ando, Journal of the Physical Society of Japan **82** (10), 102001 (2013).

[9] Z. J. Yue, C. B. Zhu, S. X. Dou, and X. L. Wang, Physical Review B **86** (19), 195120 (2012); Z. J. Yue, X. L. Wang, and S. X. Dou, Applied Physics Letters **101** (15), 152107 (2012).

[10] D. J. Kim, S. Thomas, T. Grant, J. Botimer, Z. Fisk, and J. Xia, Sci. Rep. **3** (2013).

[11] G. D. Mahan, in *Solid State Physics*, edited by Henry Ehrenreich and Spaepen Frans (Academic Press, 1997), Vol. Volume 51, pp. 81; M. Cutler and N. F. Mott, Physical Review **181** (3), 1336 (1969).

[12] J. Jiang, S. Li, T. Zhang, Z. Sun, F. Chen, Z. R. Ye, M. Xu, Q. Q. Ge, S. Y. Tan, X. H. Niu, M. Xia, B. P. Xie, Y. F. Li, X. H. Chen, H. H. Wen, and D. L. Feng, Nat Commun **4** (2013).

[13] R. Takahashi and S. Murakami, Semiconductor Science and Technology **27** (12), 124005 (2012).

[14] S.-P. Chao and V. Aji, Physical Review B **84** (15), 155430 (2011).

[15] R. Takahashi and S. Murakami, Physical Review B **81** (16), 161302 (2010).

[16] P. Ghaemi, R. S. K. Mong, and J. E. Moore, Physical Review Letters **105** (16), 166603 (2010).

[17] O. A. Tretiakov, A. Abanov, and J. Sinova, Applied Physics Letters **99** (11), 113110 (2011).

[18] O. A. Tretiakov, A. Abanov, and J. Sinova, Journal of Applied Physics **111** (7), 07E319 (2012).

[19] O. A. Tretiakov, A. Abanov, S. Murakami, and J. Sinova, Applied Physics Letters **97** (7), 073108 (2010).

## Figures

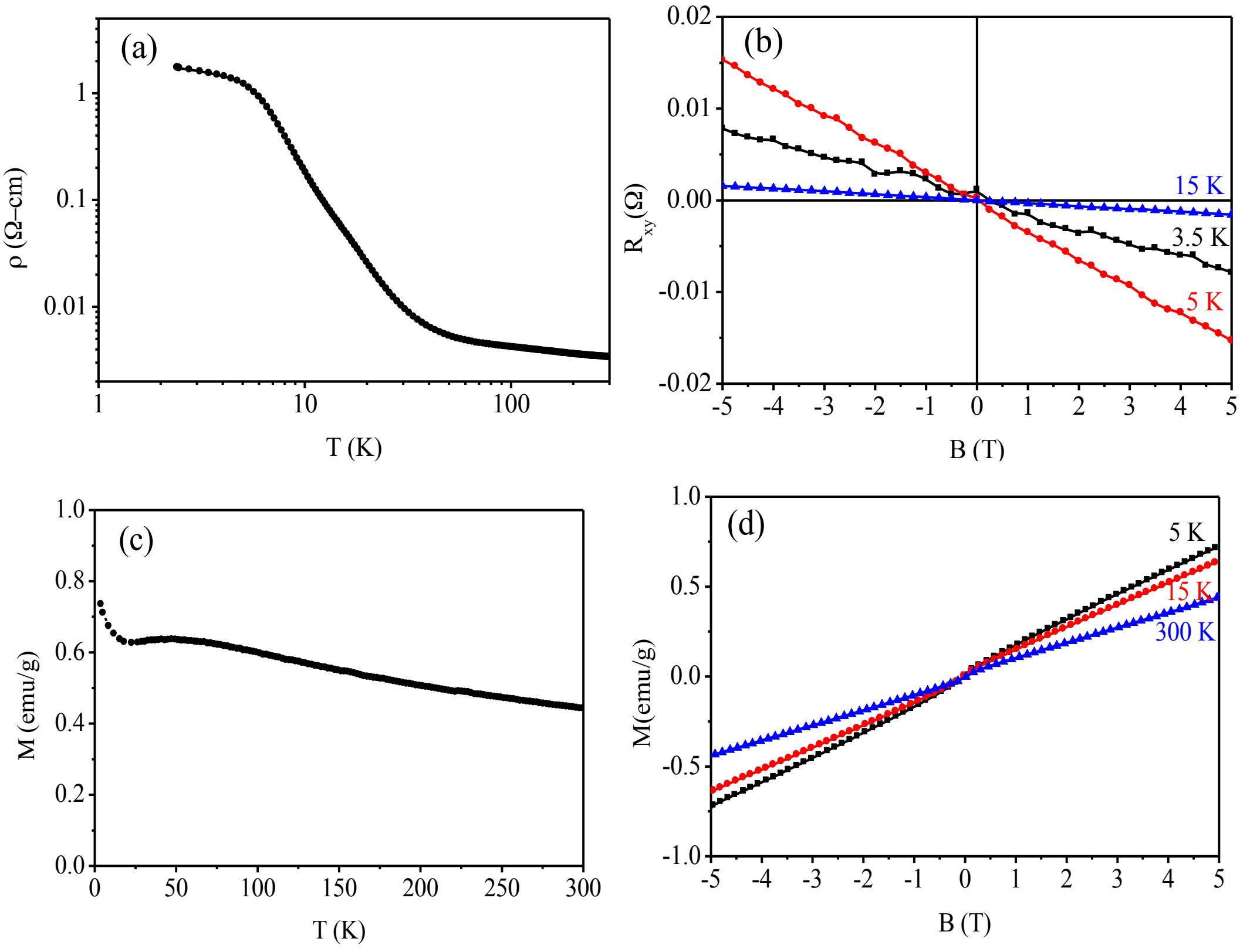

FIG. 1 (a) Resistivity of polycrystalline $SmB_6$ as function of temperature. (b) Hall resistance of polycrystalline $SmB_6$ as a function of external magnetic fields. (c) Magnetization of polycrystalline $SmB_6$ as a function of temperature. (d) Magnetization of polycrystalline $SmB_6$ as a function external magnetic fields.

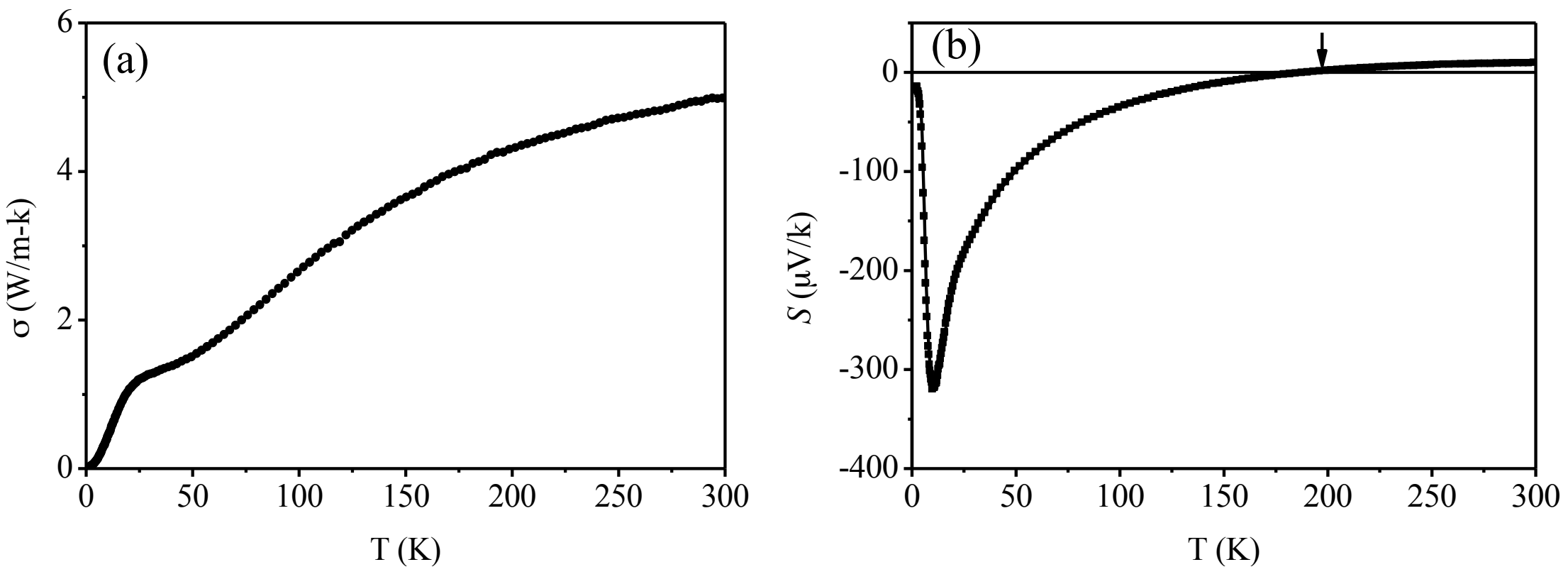

FIG. 2 (a) Thermal conductivity of polycrystalline $SmB_6$ as functions of temperature. (b) Seebeck coefficient of polycrystalline $SmB_6$ as functions of temperature.

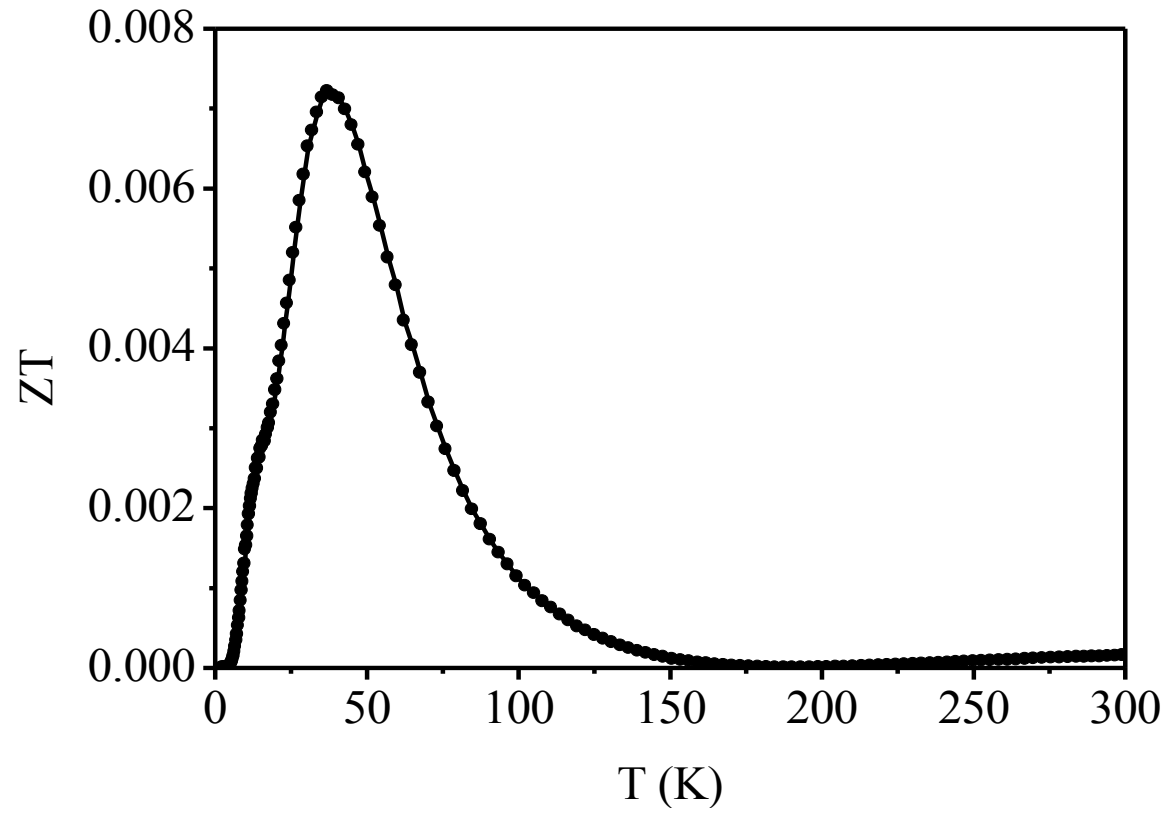

FIG. 3 *ZT* of polycrystalline $SmB_6$ as function of temperature.